\documentclass[a4paper,11pt]{article}
\usepackage{pos}
\usepackage{subcaption}

\title{Searches for and Characterization of Astrophysical Neutrinos using Starting Track Events in IceCube }
 \ShortTitle{IceCube Starting Tracks ICRC 2021}

\author{The IceCube Collaboration \\{\normalsize \normalfont(a complete list of authors can be found at the end of the proceedings)}}

\emailAdd{msilva@icecube.wisc.eduu}
\emailAdd{smancina@icecube.wisc.edu}

\abstract{
The IceCube Neutrino Observatory is a cubic kilometer-sized detector designed to detect neutrinos of astrophysical origin. However, muons created by cosmic rays interacting in the atmosphere pose a significant background for these astrophysical neutrinos particularly in the southern equatorial sky. Identifying neutrino events that start in the detector allows us to reduce the atmospheric muon component while retaining a high rate of starting neutrino events. The method presented today also rejects atmospheric neutrinos if they are accompanied by muons from the same cosmic ray shower, lowering the 50$\%$ purity threshold for astrophysical-to-atmospheric neutrinos from 100 TeV to ~10 TeV at declinations less than -25°. We use 10$\%$ (burn sample) of 9.5 years IceCube data to demonstrate the status of this dataset. We outline a planned measurement of the diffuse neutrino flux inclusive of theoretical and detector systematic uncertainties. In addition, we discuss searches for neutrino point sources and diffuse galactic plane neutrino emission in the Southern sky and plans to release high astrophysical-purity real-time alerts to the multi-messenger community.

\vspace{4mm}
{\bfseries Corresponding authors:}
Sarah Mancina$^{1}$, Manuel Silva$^{1*}$ \\
{$^{1}$ \itshape University of Wisconsin, Madison}\\[4mm]
$^*$ Presenter

\FullConference{37$^{\rm{th}}$ International Cosmic Ray Conference (ICRC 2021)\\
		July 12th -- 23rd, 2021\\
		Online -- Berlin, Germany}

}
\begin{document}
\maketitle


\section{Introduction}\label{sec:Intro}
IceCube is a cubic-kilometer neutrino detector at the geographic South Pole \cite{Aartsen:2016nxy} between depths of 1450 m and 2450 m. Reconstruction of the direction, energy and flavor of the neutrinos relies on the optical detection of Cherenkov radiation emitted by charged particles produced in the interactions of neutrinos in the surrounding ice or the nearby bedrock. Over the past decade since its construction, IceCube discovered the astrophysical neutrino flux \cite{Aartsen:2013jdh} and has measured it to $10\sigma$ \cite{Aartsen:2020aqd} significance. Since then, the diffuse flux has been measured using various techniques focusing on starting events or muon tracks from the Northern Sky \cite{Stettner:2019tok,Aartsen:2020aqd,Abbasi:2020jmh,Aartsen:2018vez}. Many searches for sources of this astrophysical flux were performed with the most significant sources to-date being the direction of TXS 0506-056 and NGC 1068 \cite{Aartsen:2019fau, IceCube:2018cha}. Despite these efforts, IceCube has yet to discover a neutrino source \cite{Aartsen:2019fau} at 5$\sigma$. 
\par
 A starting tracks dataset was published by IceCube in 2019 \cite{Aartsen:2018vez} but was statistically limited in the southern sky due to the overwhelming atmospheric muon background and the required cuts to reduce them to negligible quantities. Today, we present a new starting track event selection and it to outline a measurement of the diffuse flux and a search for neutrino sources. The starting tracks dataset takes advantage of the excellent angular and energy resolution expected with starting track events. We target events with energies above 1 TeV over the entire sky and are dominated by muon neutrinos. We briefly outline the event selection using 1 year of data to demonstrate the effectiveness of our event selection at TeV energies in Sec \ref{sec:ESTES}. A brief overview of a measurement to the astrophysical neutrino flux assuming a single power law is outlined in Sec. \ref{sec:Diffuse}. Given the high astrophysical signal purity, we then discuss the application of this dataset to search for neutrino sources in Sec. \ref{sec:NSSearch}. 


\section{Event Selection}\label{sec:ESTES}
The event selection can be summarized by two distinct cuts. The first cut, referred to as the Starting Track Veto (STV), is effective at reducing the atmospheric muon rate five orders of magnitude from ~3kHz  to ~30 mHz. The second cut uses a Boosted Decision Tree (BDT) where we tune the cut on BDT score to further reduce the muon rate to desired levels.

\begin{figure}
\centering
  \begin{subfigure}[b]{0.37\linewidth}
    \includegraphics[width=\linewidth]{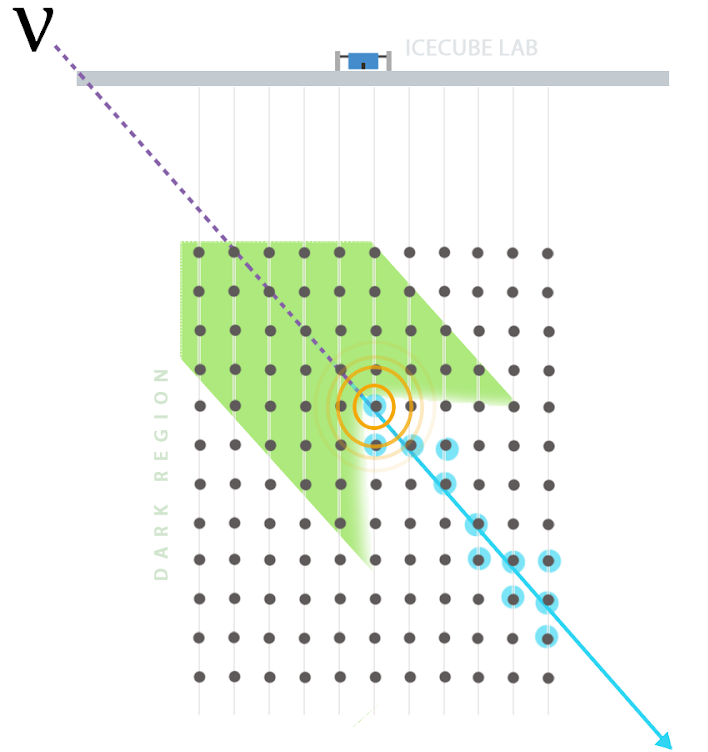}
  \end{subfigure}
  \begin{subfigure}[b]{0.385\linewidth}
    \includegraphics[width=\linewidth]{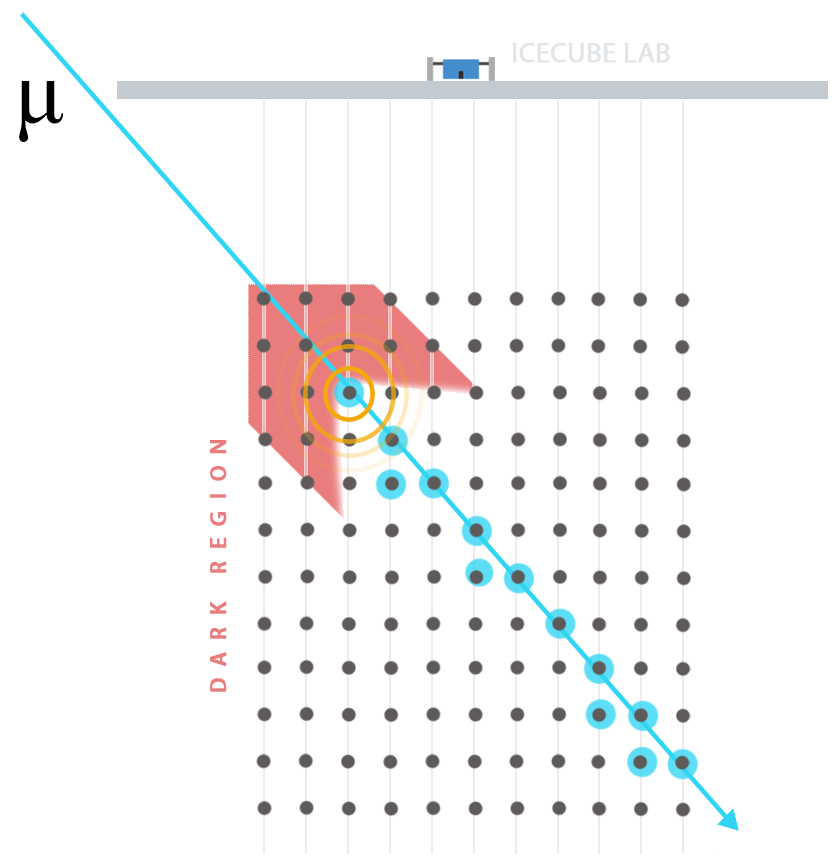}
  \end{subfigure}
  \caption{Graphic demonstrating how the starting track veto works. The orange circle is the location of the reconstructed vertex and the solid blue line represents the reconstructed muon track. No Cherenkov light is expected for an incoming neutrino before the deep inelastic collision occurs. Whereas an incoming muon deposits energy along the entire track with Cherenkov photons expected inside the dark region.}
  \label{fig:Veto_graphics}
\end{figure}

\subsection{Starting Track Veto}\label{sec:STV}
The STV uses the reconstructed track direction and vertex to define a dynamic veto region ("dark region" in Fig. \ref{fig:Veto_graphics}). We use all optical modules (OM) in the dark region to assign a score for the event's "starting" probability. This value is optimized to maximally reduce the muon rate with negligible impact on starting neutrino events. Fig.~\ref{fig:Veto_graphics} shows the dark region defined for an incoming neutrino that interacts within the volume of the detector and an atmospheric muon. The precise formulation of the STV is further discussed in Ref. \cite{Jero:2017iqd, Mancina:2019hsp, Silva:2019fnq}.

\subsection{Boosted Decision Tree Classifier}\label{sec:BDT}
\begin{figure}[h!]
\centering 
  \begin{subfigure}[b]{0.39\linewidth}
    \includegraphics[width=\linewidth]{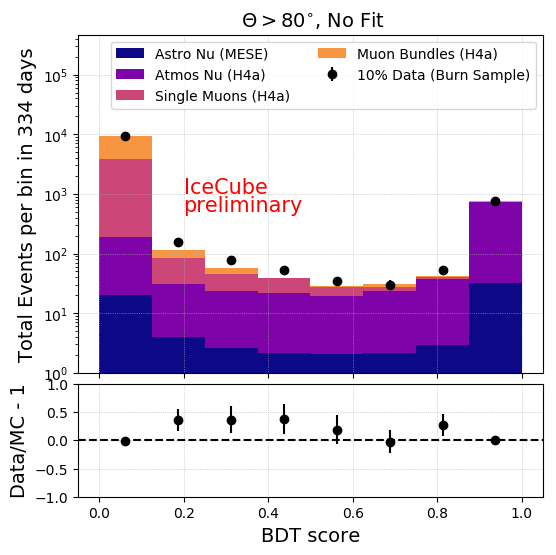}
  \end{subfigure}
  \begin{subfigure}[b]{0.39\linewidth}
    \includegraphics[width=\linewidth]{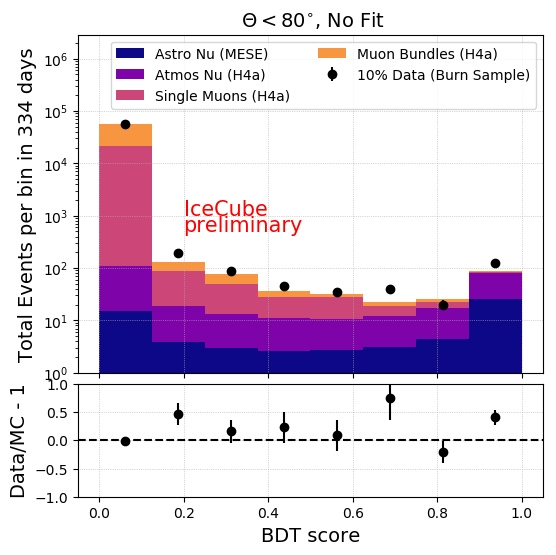}
  \end{subfigure}
  \caption{The BDT score distributions for $\Theta < 80^{\circ}$ (left) and $\Theta > 80^{\circ}$ (right). The $\Theta < 80^{\circ}$ sky is dominated by atmospheric muon events that enter between strings and deposit little light giving the illusion of being a "starting event". The $\Theta > 80^{\circ}$ sky is dominated by atmospheric muon events that arrive from the horizon or were incorrectly reconstructed in direction altogether. Due to the topological differences between the background events, two different zenith-dependent BDT scores are defined using the same trained BDT model.}
  \label{fig:bdt_scores}
\end{figure}

The XGBoost (eXtreme Gradient Boosting \cite{Chen:2016:XST:2939672.2939785}) classifier algorithm is used to reduce the atmospheric muon rate to the desired levels. We use thirteen variables to train the a classifier for atmospheric muons and starting neutrino events. We assume the 2-year flux from Ref. \cite{Aartsen:2014muf} (MESE) with spectral index = 2.46 for neutrinos and the Gaisser H4a flux \cite{Gaisser:2013bla} for muons and atmospheric neutrinos. This BDT model is then used to assign cuts as a function of zenith angle. Cuts are applied at BDT score = 0.8 for events with zenith angle $\Theta < 80^{\circ}$ (southern sky) and score = 0.9 for $\Theta > 80^{\circ}$ (northern sky). These cuts were selected to minimize the atmospheric muon rates to $\sim$handfuls per year. The BDT score distributions are shown in Fig. \ref{fig:bdt_scores} with the expected atmospheric muons separated into single muons and muon bundles for illustrative purposes. The total Monte Carlo expected rate is scaled to the total data rate to compare the BDT performance over all scores. The dataset shows good agreement between the data and simulation at high and low BDT scores. There is a large statistical uncertainty in the intermediate BDT score region, but sufficient simulation statistics are expected at the high BDT score region where the cut is eventually applied.

\subsection{Expected Rates After All Cuts}\label{sec:FinalRates}
\begin{figure}[h!]
\centering 
  \begin{subfigure}[b]{0.39\linewidth}
    \includegraphics[width=\linewidth]{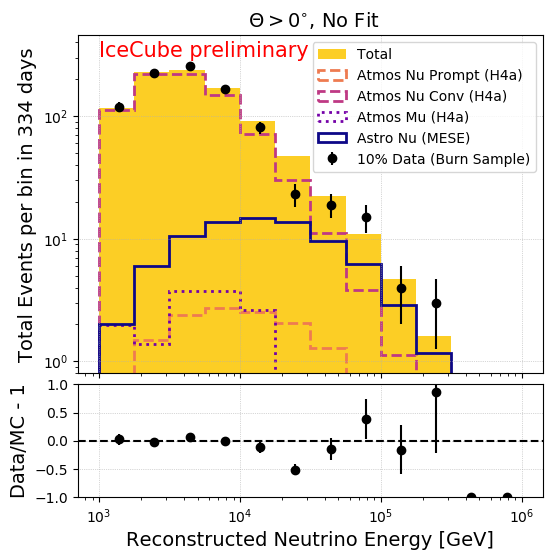}
  \end{subfigure}
    \begin{subfigure}[b]{0.39\linewidth}
    \includegraphics[width=\linewidth]{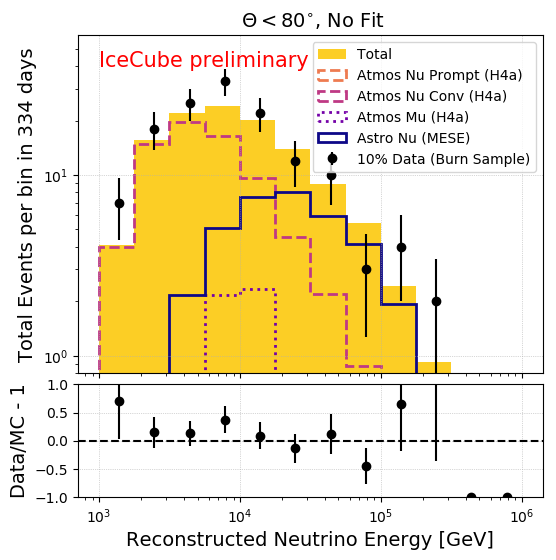}
  \end{subfigure}
  \caption{The left figure shows the dataset over the entire sky while the right figure is limited to only the southern sky. For the southern sky, we are able to detect astrophysical neutrinos with 50$\%$ purity for energies near 10 TeV.}
  \label{fig:datamc_finallevel}
\end{figure}

Fig. \ref{fig:datamc_finallevel} shows the final rates from the simulation and 334 days of data after applying the BDT cuts from section~\ref{sec:BDT}. The energy reconstruction was based off the random forest technique used in Ref. \cite{Aartsen:2018vez} with additional details found in Ref. \cite{Silva:vlvnt}. We show the dataset performance over the entire sky in Fig. \ref{fig:datamc_finallevel} (left) assuming these benchmark fluxes, Fig. \ref{fig:datamc_finallevel} (right) shows the dataset with an additional cut of $\Theta < 80^{\circ}$. Using 1 year of data and benchmark fluxes, we see there is already promising agreement with the data and simulation demonstrating that our event selection was successful in reducing the atmospheric muon background to desired levels. 

\section{Diffuse Neutrino Flux Measurement}\label{sec:Diffuse}

\subsection{Flux Models}\label{sec:Diffuse_Models}
The starting tracks dataset will consist of atmospheric muons, conventional neutrinos, prompt neutrinos, and astrophysical neutrinos. We treat the atmospheric components as nuisance parameters by fitting the a scaling factor multiplied to the expected flux to the data. Eq. \ref{eq:atmos_flux} shows how this modification is implemented into our model. We take into account the rejection of atmospheric neutrinos with accompanying muons by applying the self-veto analytical correction using NuVeto \cite{Arguelles:2018awr}. 

\begin{equation*}
  \begin{array}{l}
\Phi_{\mu} = \phi_{\mu} \times \Phi_{\mu,\text{GaisserH4a}} 
\\
\Phi_{\text{conv}\: \nu} = \phi_{\text{conv}} \times \Phi_{\text{conv,GaisserH4a}}
\\
\Phi_{\text{prompt} \:\nu} = \phi_{\text{prompt}} \times \Phi_{\text{prompt,GaisserH4a}}
\label{eq:atmos_flux}
  \end{array}
\end{equation*}

The astrophysical neutrino flux is modeled assuming a single power law flux. The astrophysical flux assumes 1:1:1 flavor ratio (this is the factor of "3" in Eq. \ref{eq:SPL}). The factor of $\Phi_{\text{astro}}$ is a unit less value that represents the per flavor astrophysical normalization at 100 TeV. The spectral index $\gamma_{\text{astro}}$ is a unit less positive definite value.

\begin{equation}
\left(\frac{\mathrm{d}\Phi}{\mathrm{dE}}\right)_{\text{Total}} = \Phi_{\text{Astro}}^{\nu+\overline\nu, \text{per flavor}} \times 3 \times (\frac{\mathrm{E}_{\nu}}{100\mathrm{TeV}})^{-\gamma_{\text{Astro}}}\; \mathrm{GeV^{-1}\cdot cm^{-2}\cdot sr^{-1}\cdot s^{-1}}
\label{eq:SPL}
\end{equation}

\subsection{Analysis Methods}\label{sec:Diffuse_Meth}
We use a forward folding binned likelihood ratio test to perform our measurement of the diffuse flux. All events with reconstructed energy above 1 TeV and below 1 PeV over the entire sky are used. We define 10 bins in energy and 8 bins in zenith, where each i-th bin contains a Monte Carlo expectation ($\lambda_{i}$) and observed data ($k_{i}$). Each bin is modeled assuming a Poisson probability with our log-likelihood defined as the sum over all bins outlined in Eq. \ref{eq:llh}.

\begin{equation}
\text{log} \mathcal{L} (\lambda,x) = \Sigma _{i=1}^{80} (x_{i}\text{log}\lambda_i - \lambda_i)  +  \sum^{syst}_{j} \frac{(x_j-\mu_j)^2}{\sigma_j^2}
\label{eq:llh}
 \end{equation}

We include additional nuisance parameters from detector and theoretical uncertainties in the likelihood as a Gaussian with priors. We use MCEq \cite{Gaisser:2019xlw} and NuVeto to calculate the expected atmospheric flux using various cosmic ray \cite{Gaisser:2013bla,Hoerandel:2002yg,Zatsepin:2006ci} and hadronic interaction\cite{Ostapchenko:2010vb,Pierog:2013ria,Riehn:2017mfm} models. The shape differences between the various models are interpolated and used to define the theoretical uncertainties \cite{Silva:vlvnt}. The bulk ice absorption length coefficients \cite{Ackermann:2006pva} and overall detector response are each independently varied by $\pm10\%$. There are two angular dependent efficiency factors used to model the detector's response to the hole-ice where the strings were placed.

\subsection{Measurement of the Astrophysical Diffuse Flux}\label{sec:Diffuse_Obs}
\begin{figure}[h!]
\centering 
  \begin{subfigure}[b]{0.85\linewidth}
    \includegraphics[width=\linewidth]{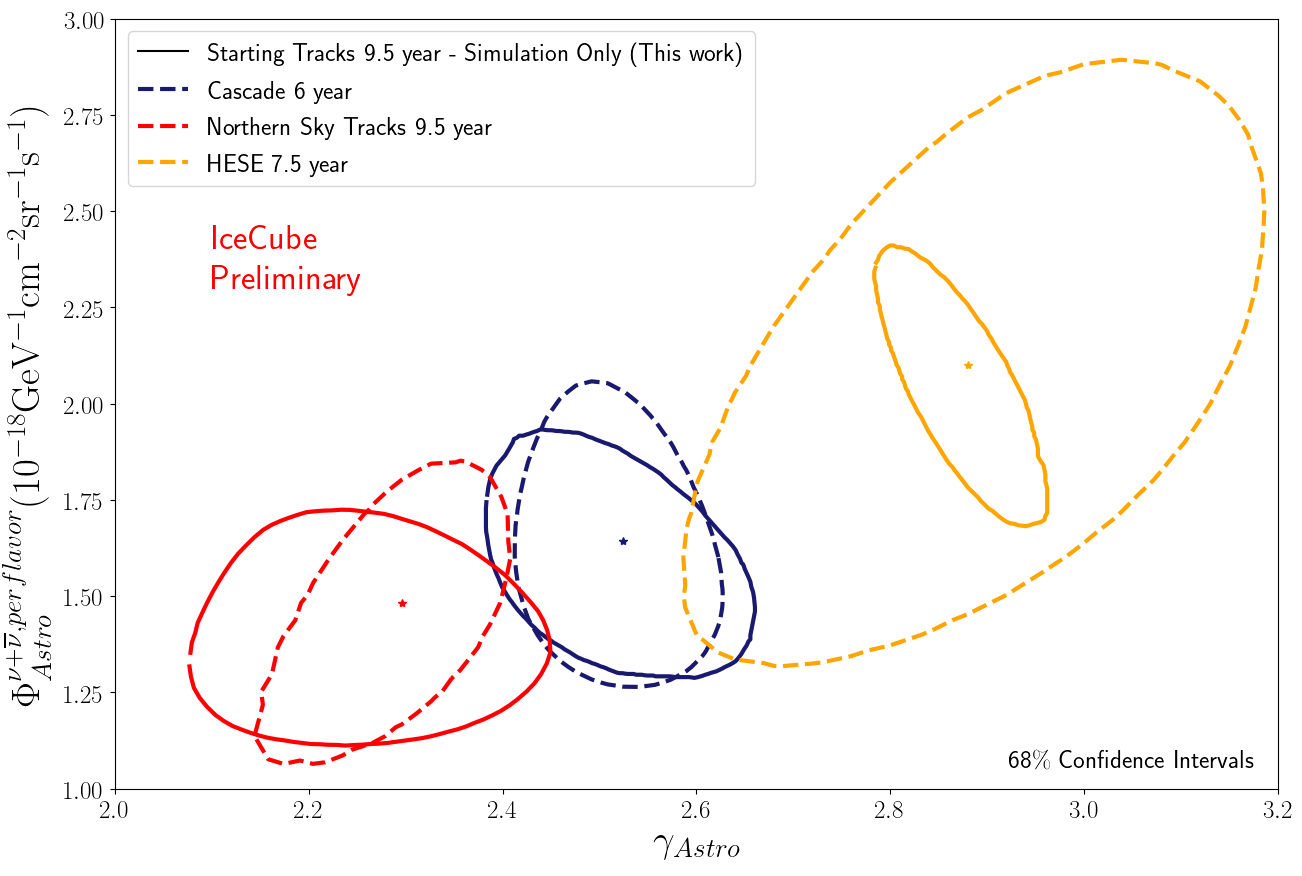}
  \end{subfigure}
  \caption{The single power law likelihood scans with the best fit point shown as a star. The confidence intervals for this sensitivity measurement show the expected precision with this dataset assuming 9.5 years of data. The CI are reduced in size as the spectrum becomes softer and as the normalization increase as a result of the expected increase in astrophysical neutrino rates at TeV energies.}
  \label{fig:spl_fit}
\end{figure}

We assume 9.5 years of IceCube-86 data will be used for this measurement. Expected 1$\sigma$ sensitivity to the astrophysical flux is shown in Fig.~\ref{fig:spl_fit} assuming the best-fit fluxes from the Northern Tracks 10 year\cite{Stettner:2019tok}, Cascade 6 year\cite{Aartsen:2020aqd}, and HESE 7.5 year \cite{Abbasi:2020jmh} measurements. For harder fluxes, the expected neutrino rates are low leading to large contours in both $\Phi_{\text{Astro}}$ and $\gamma_{\text{Astro}}$. For softer fluxes, the expected neutrino rates are very large leading to significantly reduced contour sizes. However, the uncertainty in $\Phi_{\text{Astro}}$ is larger than $\gamma_{\text{Astro}}$ due to the increased degeneracy with the $\phi_{\text{conv}}$ and $\phi_{\text{pr}}$ fluxes. The $\phi_{\mu}$ components contributes largely to the uncertainty of the spectral index measurement but will be improved with additional cuts in the near future.

\section{Search for Neutrino Sources}\label{sec:NSSearch}
\begin{figure}[h!]
\centering 
  \begin{subfigure}[b]{0.65\linewidth}
    \includegraphics[width=\linewidth]{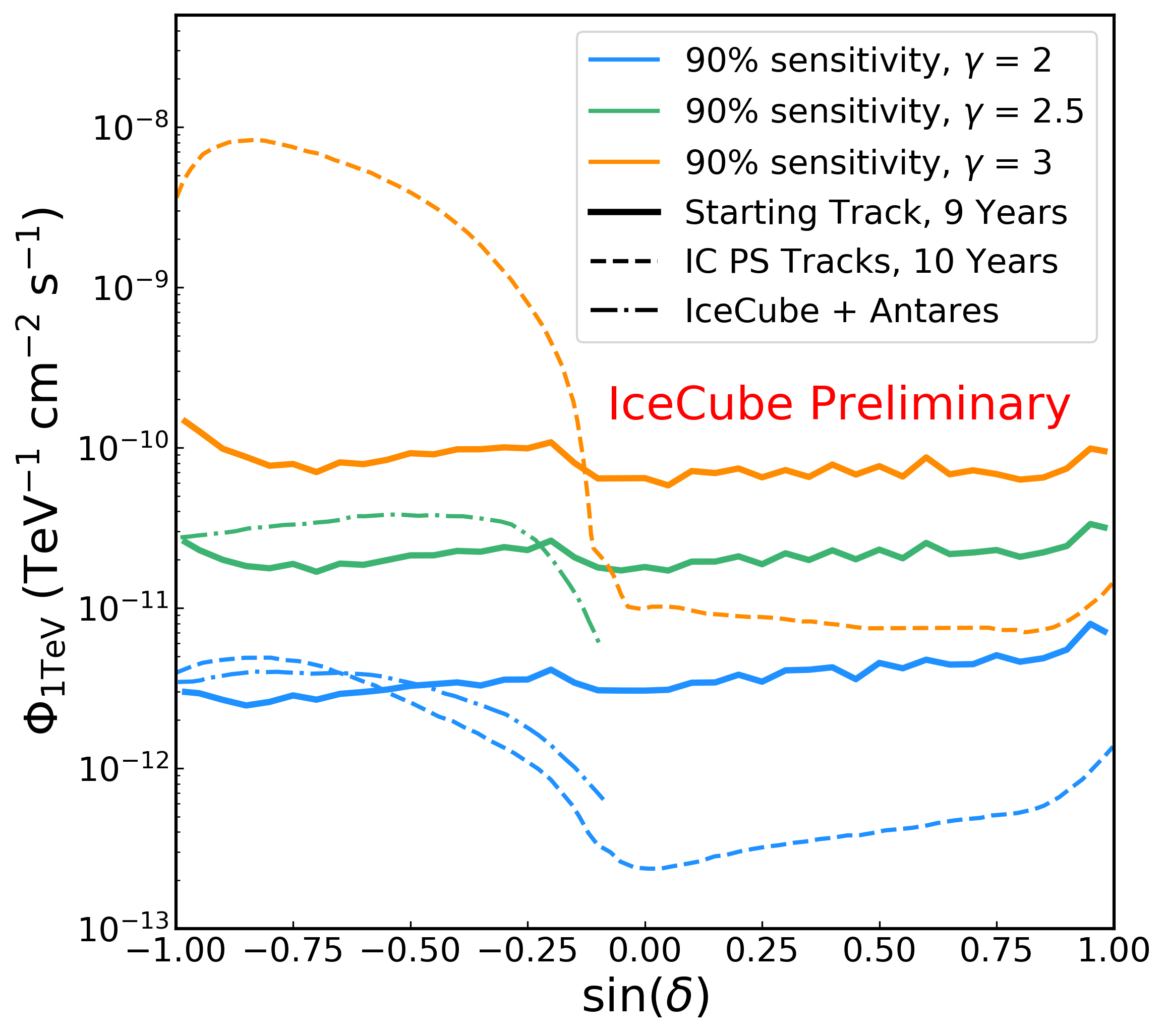}
  \end{subfigure}
  \caption{The 90$\%$ sensitivity fluxes computed at 1 TeV calculated using the technique from Ref. \cite{Abbasi_2011}. The starting tracks dataset is most competitive in the southern sky and will contribute greatly to understanding what processes produce astrophysical neutrinos.}
  \label{fig:all_sky}
\end{figure}

We now outline a set of proposed searches for neutrino sources. We plan to perform a search for neutrinos over the entire sky treating the spectral index as a free parameter using the techniques outlined in Ref.~\cite{Abbasi_2011}. The sensitivities in Fig.~\ref{fig:all_sky} demonstrate our performance over all declination compared to the IceCube 10 year time-integrated search \cite{Aartsen:2019fau} and the combined IC+Antares search \cite{Aartsen:2020xpf}. For softer sources, this result is expected to improve the sensitivity by 2 orders of magnitude mostly attributed to the high astrophysical purity towards 10 TeV neutrino energies. For harder sources, our expected sensitivity is a $\approx 50 \%$ improvement for declinations less than $-25^{\circ}$. The overall improvements shown by this dataset are driven by the lower energy neutrino events, whereas previous efforts relied on high energy selection criteria resulting in significantly reduced rates in the southern sky.

Given the excellent purity of astrophysical neutrinos at 10 TeV, we plan to release real time alerts with a selection similar to what was presented here. Integrating over energies above 1 TeV and zenith angles less than 90$^\circ$, we estimate that there will be 6.7 alerts per year with at least 50$\%$ astrophysical purity.


\section{Summary and Discussion}\label{sec:Conclusion}
We outlined a new event selection to identify starting track events in IceCube. Using benchmark atmospheric and astrophysical flux measurements from recent IceCube measurements, we showed that there is good agreement with the data and the Monte Carlo. We outlined a proposed measurement of the diffuse neutrino flux and showed that the precision is competitive with recent measurements with a unique event topology. This dataset was then used to estimate our sensitivity to search for sources of neutrinos over the entire sky, showing that we expect to be competitive at declinations less than -25°.

The treatment of the systematic uncertainties is close to completion with a measurement of the diffuse neutrino flux to follow. In parallel to these efforts, we are preparing to perform our search for neutrino clustering over the entire sky. These results will be followed by a real-time alert stream to the multi-messenger community. The details of these searches are in development and approaching release.


\clearpage
\section*{Full Author List: IceCube Collaboration}

\scriptsize
\noindent
R. Abbasi$^{17}$,
M. Ackermann$^{59}$,
J. Adams$^{18}$,
J. A. Aguilar$^{12}$,
M. Ahlers$^{22}$,
M. Ahrens$^{50}$,
C. Alispach$^{28}$,
A. A. Alves Jr.$^{31}$,
N. M. Amin$^{42}$,
R. An$^{14}$
% ...
\\

\noindent
% ...
$^{12}$ Universit{\'e} Libre de Bruxelles, Science Faculty CP230, B-1050 Brussels, Belgium \\
% ...
$^{14}$ Department of Physics and Laboratory for Particle Physics and Cosmology, Harvard University, Cambridge, MA 02138, USA \\
% ...
$^{17}$ Department of Physics, Loyola University Chicago, Chicago, IL 60660, USA \\
$^{18}$ Dept. of Physics and Astronomy, University of Canterbury, Private Bag 4800, Christchurch, New Zealand \\
% ...
$^{22}$ Niels Bohr Institute, University of Copenhagen, DK-2100 Copenhagen, Denmark \\
% ...
$^{28}$ D{\'e}partement de physique nucl{\'e}aire et corpusculaire, Universit{\'e} de Gen{\`e}ve, CH-1211 Gen{\`e}ve, Switzerland \\
% ...
$^{31}$ Karlsruhe Institute of Technology, Institute for Astroparticle Physics, D-76021 Karlsruhe, Germany  \\
% ...
$^{42}$ Bartol Research Institute and Dept. of Physics and Astronomy, University of Delaware, Newark, DE 19716, USA \\
% ...
$^{50}$ Oskar Klein Centre and Dept. of Physics, Stockholm University, SE-10691 Stockholm, Sweden \\
% ...
$^{59}$ DESY, D-15738 Zeuthen, Germany \\
% ...

\subsection*{Acknowledgements}

\noindent
USA {\textendash} U.S. National Science Foundation-Office of Polar Programs,
U.S. National Science Foundation-Physics Division,
U.S. National Science Foundation-EPSCoR,
Wisconsin Alumni Research Foundation,
Center for High Throughput Computing (CHTC) at the University of Wisconsin{\textendash}Madison,
Open Science Grid (OSG),
Extreme Science and Engineering Discovery Environment (XSEDE),
Frontera computing project at the Texas Advanced Computing Center,
U.S. Department of Energy-National Energy Research Scientific Computing Center,
Particle astrophysics research computing center at the University of Maryland,
Institute for Cyber-Enabled Research at Michigan State University,
and Astroparticle physics computational facility at Marquette University;
Belgium {\textendash} Funds for Scientific Research (FRS-FNRS and FWO),
FWO Odysseus and Big Science programmes,
and Belgian Federal Science Policy Office (Belspo);
Germany {\textendash} Bundesministerium f{\"u}r Bildung und Forschung (BMBF),
Deutsche Forschungsgemeinschaft (DFG),
Helmholtz Alliance for Astroparticle Physics (HAP),
Initiative and Networking Fund of the Helmholtz Association,
Deutsches Elektronen Synchrotron (DESY),
and High Performance Computing cluster of the RWTH Aachen;
Sweden {\textendash} Swedish Research Council,
Swedish Polar Research Secretariat,
Swedish National Infrastructure for Computing (SNIC),
and Knut and Alice Wallenberg Foundation;
Australia {\textendash} Australian Research Council;
Canada {\textendash} Natural Sciences and Engineering Research Council of Canada,
Calcul Qu{\'e}bec, Compute Ontario, Canada Foundation for Innovation, WestGrid, and Compute Canada;
Denmark {\textendash} Villum Fonden and Carlsberg Foundation;
New Zealand {\textendash} Marsden Fund;
Japan {\textendash} Japan Society for Promotion of Science (JSPS)
and Institute for Global Prominent Research (IGPR) of Chiba University;
Korea {\textendash} National Research Foundation of Korea (NRF);
Switzerland {\textendash} Swiss National Science Foundation (SNSF);
United Kingdom {\textendash} Department of Physics, University of Oxford.

\end{document}